\documentclass[letterpaper]{article} % DO NOT CHANGE THIS
\usepackage{aaai2026}  % DO NOT CHANGE THIS
\usepackage{times}  % DO NOT CHANGE THIS
\usepackage{helvet}  % DO NOT CHANGE THIS
\usepackage{courier}  % DO NOT CHANGE THIS
\usepackage[hyphens]{url}  % DO NOT CHANGE THIS
\usepackage{graphicx} % DO NOT CHANGE THIS
\usepackage{natbib}  % DO NOT CHANGE THIS AND DO NOT ADD ANY OPTIONS TO IT
\usepackage{caption} % DO NOT CHANGE THIS AND DO NOT ADD ANY OPTIONS TO IT
\usepackage{algorithm}
\usepackage{algorithmic}
\usepackage{float}

\usepackage{newfloat}
\usepackage{listings}
\DeclareCaptionStyle{ruled}{labelfont=normalfont,labelsep=colon,strut=off} % DO NOT CHANGE THIS
\floatstyle{ruled}
\newfloat{listing}{tb}{lst}{}
\floatname{listing}{Listing}
\title{Appearing Legitimate is Not Enough: Interrogating Synthetic Agents in Representational Processes through a Participatory Design Lens}

\author{
    Aditya Nayak\equalcontrib\textsuperscript{\rm 1},
    Aditi Vashistha\equalcontrib\textsuperscript{\rm 2},
    Alissa Centivany\textsuperscript{\rm 2},
    Aakash Gautam\textsuperscript{\rm 1}
}
\affiliations{
    \textsuperscript{\rm 1}University of Pittsburgh\\
    \textsuperscript{\rm 2}FIMS, Western University\\
    aditya.nayak@pitt.edu, avashi4@uwo.ca, acentiva@uwo.ca, aakash@pitt.edu
}

\begin{document}

\maketitle

\begin{abstract}

Synthetic agents built atop LLM-based foundation models are gaining popularity as substitutes for human participants across research contexts, including user-testing, market-research, computational social science, surveys, and qualitative research. We are also witnessing an extension of synthetic agents into experimental implementations of policy consultation, jury deliberation, humanitarian diplomacy, and similar contexts where human participation and representation are central to the perceived legitimacy of the institutional processes. The value of participation extends beyond informational contributions and consensus generation; participation is a necessary, legitimizing condition for democratic political institutions and processes. Treating synthetic agents as human substitutes raises serious political, representational, and ethical concerns. Participatory Design’s modes of engagement — probing, priming, understanding, and generating — offer helpful tools for engaging with representational questions of personhood. We apply the lens to three case studies of synthetic agents substituting for personhood at varying representational scales: local policy, enterprise jury deliberation, and global diplomacy. We argue that legitimacy and personhood are integral and mutually constitutive while identifying the ethical, representational, and methodological risks of using synthetic agents in representational processes. We conclude by proposing soft and hard boundaries for designing oversight on LLMs and synthetic agents in representational processes.

\end{abstract}

\section{Introduction}

Large Language Models and agents built on top of the foundational models are increasingly being deployed as substitutes for human participants in domains like user-testing \cite{yang_carrier_2026}, market-research \cite{simileai_home_nodate}, computational social science \cite{wang_large_2025}, surveys \cite{ball_human_2025}, and qualitative research \cite{nguyen_generative_2026}. 
Synthetic agents, that is, LLM-based systems trained or augmented with domain-specific data to mimic context-bound personas, are now being extended into experimental implementations of policy consultation \cite{reshef_kera_governance_2024}, jury deliberation \cite{ersoz_12_2026}, and humanitarian diplomacy \cite{albrecht_does_2025}.
Their proponents claim these systems can perceive contextual environments, make decisions, and take actions in ways that approximate human decision-making.

% Synthetic agents are LLM-based systems that are trained on a specific selection of data, often domain-specific, to add on to the generalized training of the foundational models. This gives synthetic agents a characteristic form, and they can be claimed to draw on context-specific data for acting in the domain. The proponents claim that synthetic agents can perceive their contextual environment, make decisions, and take actions, in effect mimicking human-like decision-making and problem-solving. 

The use of synthetic agents is often justified as a solution to the difficulty of recruiting suitable participant groups.
One could argue that in low-risk domains, like market research, the stakes of this substitution are limited to participant representativeness and data quality \cite{maden_three_2026, su_toward_2026} for building better products. 
While that in itself is contentious, the use of synthetic agents in domains like policy making, diplomatic consultation, and legal juries is alarming.  
These are deliberative and representational processes where participation and representation extend beyond informational contribution and consensus generation. We use representational process to mean a deliberative or decision-making process whose legitimacy is procedural. It derives from the inclusion and participation of those affected by the decision instead of the correctness of the outcome \cite{habermas_between_1996, young_inclusion_2000}.
Participation is a necessary and legitimizing condition of the democratic process, and importantly, it is fundamental to the institutional legitimacy these processes claim. 
Substituting synthetic agents for human participants in these contexts raises political, representational, and ethical concerns, including the exclusion of the lived experiences of the marginalized groups who would otherwise have been participants.

The concept of personhood has been made and remade whenever societies need new vocabulary for new circumstances \cite{leibo_pragmatic_2025}. 
But most conceptions of personhood in moral and legal domains rest on consciousness, intentionality, the capacity for reciprocal moral relations, and responsibility and accountability for the consequences of one's actions \cite{baeyaert_beyond_2025, bronner-martin_ethical_2025}. 
Synthetic agents do not possess these properties, and prior work argues that this limitation cannot be overcome through technical fixes \cite{bronner-martin_ethical_2025}. 
Despite this, synthetic agents are being deployed in roles that presuppose them.

% Existing literature on simulated research participation has highlighted issues with the use of LLMs. These includes concerns around positionality \cite{wang_large_2025}, validity \cite{nguyen_generative_2026,agnew_workshop_2026,maden_three_2026,jowsey_we_2025}, epistemic risks \cite{nguyen_generative_2026}, and ethical implications \cite{agnew_workshop_2026} such as retrospective consent and moral alignment of models. But most of the existing research explores these problems in the context of user research. Most frameworks of evaluation of these models are outcome-based including developing frameworks for 'Believability' \cite{xiao_how_2024}, 'Faithfulness' \cite{maden_three_2026, agarwal_faithfulness_2024}, 'Plausibility' \cite{agarwal_faithfulness_2024}, 'Algorithmic Fidelity' \cite{argyle_out_2023}. Given the sensitive and high-stakes nature of policy, juries, and diplomacy, it is important to evaluate the application of synthetic agents in these domains.

%% What is our secret sauce?
In this paper, we investigate how synthetic agents come to appear legitimate as substitutes for human participants in representational processes. 
We approach this through three case studies at different scales, namely local policy (Ana, a community-voice chatbot), enterprise legal deliberation (Synthetic Juror), and global diplomacy (Ask Amina and Ask Abdalla).
To analyze these cases, we draw on Sanders et al.'s framework of Participatory Design (PD), which identifies four modes through which a participant is produced from a person, namely, \emph{probing, priming, understanding,} and \emph{generating} \cite{sanders_framework_2010}.
These modes give us process-oriented perspectives to evaluate what representational work the synthetic agents are doing, and what they leave out.
We ask two research questions: % save space by listing them in the same paragraph.
\textbf{RQ1}: How do synthetic agents in representational processes construct an appearance of legitimate participation?\
\textbf{RQ2}: What boundaries should guide the design and deployment of synthetic agents in representational processes?

% The complexities of representational process are traditionally handled by employing several modes of engagement. Participatory Design (PD) is one such domain that attempts to involve participants in the design process in various capacities. PD utilizes different modes of engagement for this purpose. Sanders et al. \cite{sanders_framework_2010} have provided a framework to think about the tools of engagement in PD.

Across the three cases, we trace a four-step process through which personhood is manufactured. % Add rich points to each if you can. 
First, through the framing of the problem. 
Then, through the institutionalized curation of data.
This is followed by the design of user encounters with the persona. 
Finally, through the technical evaluation of validity. 
We argue that these four steps collectively produce \emph{manufactured personhood}, an appearance of legitimate participation that bypasses the representational processes from which institutional legitimacy is derived. 
We contrast our position against existing evaluation frameworks for synthetic agents, such as \textit{believability} \cite{xiao_how_2024}, \textit{faithfulness} \cite{maden_three_2026, agarwal_faithfulness_2024}, \textit{plausibility} \cite{agarwal_faithfulness_2024}, and \textit{algorithmic fidelity} \cite{argyle_out_2023}, all of which cannot detect this bypass because they measure similarity of outputs rather than the integrity of the underlying processes.

With this paper, we make three contributions. 
First, by examining three case studies, we offer a four-step account of how synthetic agents manufacture personhood across representational scales.
Second, we use the PD modes of engagement to develop a process-oriented critique of synthetic agents, complementing the outcome-centric evaluations that dominate the existing literature. 
Third, we propose desiderata for design consisting of \emph{soft boundaries} that separate analytic and indexing uses of LLMs from anthropomorphized interfaces, and \emph{hard boundaries} that categorically separate anthropomorphized interfaces from systems that substitute for human participants in representational processes.
%% Why now?
We make these proposals at a moment when synthetic agents in representational contexts remain experimental, and the boundaries are still tractable.

\section{Related Works}

%Large language models are becoming increasingly popular in their application as a replacement of human participation in domains like user-testing \cite{yang_carrier_2026}, computational social science \cite{wang_large_2025}, surveys \cite{ball_human_2025}, and qualitative research \cite{nguyen_generative_2026}. There are also recent experimental studies exploring the application of synthetic agents in policy deliberations \cite{reshef_kera_governance_2024}, jury trials \cite{ersoz_12_2026}, and humanitarian diplomacy \cite{albrecht_does_2025}.

\subsection{Synthetic Agents as Research Participants}

The use of synthetic agents as substitutes for human participants raises a set of interrelated concerns. 
These include positionality, validity, epistemic and ethical risks, and the boundaries of knowledge production \cite{alvarado_garcia_disciplinary_2025, wang_large_2025, nguyen_generative_2026, agnew_workshop_2026, jowsey_we_2025, maden_three_2026}.
Scholars have cautioned against using synthetic agents as wholesale replacements for human participants \cite{arawjo_thats_2026}, and a recent ACM workshop \cite{agnew_workshop_2026} has begun to consolidate standards and documentation practices for LLM use as simulated research participants \cite{su_toward_2026, kuric_synthetic_2026, kim_layer-specific_2026, sampson_surveying_2026, maden_three_2026, yang_carrier_2026}. %% Fix this sentence: What are these citations for? Are these the standards that have been consolidated?

A recurring critique concerns how synthetic agents handle social identity. 
To represent humans, LLMs would need to capture the influence of positionality, including the relevance of social identities like race and gender \cite{wang_large_2025}.
In practice, they fail, for example, with misportrayal and category error \cite{wang_large_2025, nguyen_generative_2026}, group flattening, and identity essentialization \cite{su_toward_2026}.
Since these systems are trained on data dominated by majority populations, they gravitate toward median experiences and reject tail cases, marginalizing the populations who already are disproportionately and negatively impacted by technology \cite{su_toward_2026, smith_crafting_2025}, and producing cultural scarcity \cite{alvarado_garcia_disciplinary_2025}. %%That UCLA prof's book too
Models reduce complex demographic dynamics, such as systemic inequality produced through discriminatory policies, historical neglect, or economic exclusion, to mere demographic proxies \cite{maden_three_2026}. 
Simulation results can appear believable on the surface while the underlying logic, that of treating identity as a collection of inherent traits, is inaccurate \cite{maden_three_2026, su_toward_2026}.

%Results of simulation can appear 'believable' on surface but the logic behind it is inaccurate\cite{maden_three_2026}. Models treat identity as a collection of inherent traits rather than recognizing something shaped by external forces \cite{su_toward_2026}. 

A second strand of critique concerns embodiment. 
Synthetic agents lack the embodied lived experiences that human participants bring to research \cite{agnew_workshop_2026}, and in spatial usability evaluation they perform poorly on embodied experience criteria \cite{kim_layer-specific_2026}.
Their implementations often treat surface-level markers of human realism as evidence of behavioral or experiential validity \cite{su_toward_2026}.

Problems in evaluation follow naturally from these concerns. 
Human participation evaluation centers on identity verification grounded in lived experience. 
LLMs have no lived experience to verify, and the task shifts from verification to characterization, such as `does this simulated entity faithfully reflect properties of the population the researcher intends to study?' 
With human participants, data quality is assessed through depth, richness, and experiential specificity, none of which can be automatically measured \cite{maden_three_2026}.
In this space, scholars have proposed alternative evaluation frameworks: \textit{believability}, defined through metrics of consistency with human-like output and robustness under disturbance \cite{xiao_how_2024}; \textit{faithfulness} \cite{maden_three_2026, agarwal_faithfulness_2024}; \textit{plausibility} \cite{agarwal_faithfulness_2024}; and \textit{algorithmic fidelity} \cite{argyle_out_2023}.
However, these frameworks are all output-centric. 
They do not examine the processes that lead to those outputs. 
Beyond evaluation, prior work identifies ethical concerns around retrospective consent and the moral alignment of models \cite{agnew_workshop_2026}, and epistemic risks including category error, misattribution, the \textit{oracle effect}, and anthropomorphic fallacies \cite{nguyen_generative_2026, stark_animation_2024, maeda_anthropomorphism_2025, maeda_when_2024}.
%This body of work largely concerns user research and qualitative research contexts.
While some work examine legal implications of agents in e-commerce \cite{riedl_ai_2025}, and socio-political implications of ``agency-washing'' \cite{timaite_agents_2025}, empirical studies of application of synthetic agents as participants in representational processes remain underexplored, and importantly, the high stakes of policy, juries, and diplomacy make this gap consequential.

%Prior works also highlighted the ethical concerns \cite{agnew_workshop_2026,su_toward_2026} and risks \cite{nguyen_generative_2026} emerging out of using synthetic agents as replacement of human participants. The major ethical concerns involve retrospective consent and model's moral alignment \cite{agnew_workshop_2026}. The risks associated with use of synthetic agents include epistemic risks such as category error, misattribution, and 'oracle effect' \cite{nguyen_generative_2026}. Additionally, it also risks anthropomorphic fallacies \cite{nguyen_generative_2026,stark_animation_2024}. 

%In response to this lacuna, scholars have proposed alternative evaluation frameworks like 'Believability' (defined on the basis of metrics of '\textit{consistency}' with human like output and '\textit{robustness}' of LLMs when faced with disturbances \cite{xiao_how_2024}), 'Faithfulness' \cite{maden_three_2026,agarwal_faithfulness_2024}, 'Plausibility' \cite{agarwal_faithfulness_2024}, and 'Algorithmic Fidelity' \cite{argyle_out_2023}. All of these frameworks are output centric evaluations. The field currently lacks process oriented evaluation frameworks. 

\subsection{Personhood and Legitimacy}

The English word ``\textit{person}'' is derived from the Latin word persona which refers to the mask worn by an actor portraying a character within the context of a play \cite{gunkel_debate_2021}. 
%There are different approaches to look at personhood. 
There is no fixed concept of personhood; it is made and remade whenever societies need vocabulary for new circumstances \cite{leibo_pragmatic_2025}.

Legal literature distinguishes natural and legal persons. 
A natural person has intrinsic qualities like consciousness \cite{leibo_pragmatic_2025} and is natural by virtue of being born. 
% Humans are called natural persons because they are persons by the virtue of being born, not by legal decree. 
This personhood follows from birth, not from legal decree.
Corporations and environmental entities are legal persons by decree, artificial and non-natural \cite{gunkel_debate_2021, leibo_pragmatic_2025, baeyaert_beyond_2025}. 
Legal personhood rests on recognition as a subject of law, possessing both rights and responsibilities \cite{gunkel_debate_2021}, and is historically a pragmatic instrument of governance rather than a recognition of moral worth.
Its extension to non-human entities is justified on three grounds \cite{baeyaert_beyond_2025}: a \textit{rights-based} model grounded in moral subjectivity (self-awareness, sentience, the capacity to experience harm, the ability to participate in reciprocal moral relations); a \textit{functional} model where legal status is extended as a juridical tool for coherence, welfare, or governance efficacy; and an \textit{agency-based} model grounded in intentionality, goal formation, and bearing responsibility for the consequences of action.

Across all three grounds, AI fits poorly. 
Even artificial entities like corporations are embedded in networks of human control, whereas AI systems can produce outcomes unforeseen by their developers \cite{mayne_llms_2025}. 
Moral action, in particular, requires consciousness, intentionality, and the capacity to experience and take responsibility for the consequences.
Attempts to evaluate LLMs in domains requiring moral reasoning have found their moral judgment shallow and misaligned with human reasoning \cite{jiang_can_2022, chakraborty_structured_2025}.
Prior work argues that the question of whether AI can be made more ethical is categorically misplaced since the inability of AI to be a moral actor is a philosophical limit and cannot be overcome by technical fixes \cite{bronner-martin_ethical_2025}. The capacity for reciprocal moral relations follows Strawson’s account of the reactive attitudes that constitute a moral community \cite{Strawson1962-STRFAR}. Synthetic agents can be objects of such attitudes but never subjects of them. Moreover, the technical conceptions of agents do not map to socio-legal conceptions of agency \cite{riedl_ai_2025}. Conceptions from sociology and linguistic anthropology  identify agency as ontologically relational, and inseparable from social accountability \cite{timaite_agents_2025}.
A `moral actor' in AI ethics is defined as a person who has a `will to action' \cite{laflamme_including_2023}, and a legitimate actor is one whose agency, intentionality, and capacity for reciprocal moral relationships make them bound by moral responsibility and its consequences \cite{baeyaert_beyond_2025}.
Synthetic agents do not meet this bar. 

%% Figure 5 can go here but only if we have extra space.

% \begin{figure}[h!]
%     \centering
%     \includegraphics[width=0.90\linewidth]{Paper_Content/Personhood.png}
%     \caption{Dimensions constituting personhood.}
%     \label{fig:Personhood}
% \end{figure}

\subsection{Personhood in Participatory Design}

Representation of personhood is the central problem in participatory design (PD). 
%The complexities of the representational process are engaged in PD with using several tools and techniques.  
Sanders et al.'s \cite{sanders_framework_2010} framework categorizes the purpose of design engagement into four categories: probing, priming (immersion in the domain of interest), understanding (of current experience) and generating (creating and exploring future scenarios). 
Each mode aligns with other aspects of participation, including form (making/telling/enacting) and context (stakeholder relationships, online or in-person modalities, venue, group size, and composition) \cite{sanders_framework_2010}.
Crucially, participatory modes of engagement produce a \emph{participant} from a person. 
The participant is constituted in the process of participation, but never represents the person as a whole. 
The production of a participant is mediated through the modes of engagement, each in different capacities.

\textbf{Probing}: The foundational work, `Cultural Probes' \cite{Gaver_cultural_1999} introduced probing as a participatory mode. 
Probing explores `functions', `experiences', and `cultural placement' outside the norm to discover new pleasures, forms of sociability, and cultural forms \cite{Gaver_cultural_1999, burrows_cultural_2015, madden_probes_2014}. 
It uses provocations directed towards `inspirational data' rather than `information' with the purpose of stimulating imagination and asks the question `What inspires a person's experience of a phenomenon of interest?' (e.g., `What inspires fear of vaccines in a population?'), effectively producing the \textit{a priori} of a participant in research.

\textbf{Priming}: Priming is commonly used in behavioral psychology to investigate human judgments and decision-making. 
It affects perceptions and behaviors by increasing cognitive accessibility of specific mental content. 
In the context of design, priming is achieved by using stimuli, such as drawings, videos, neutral or hostile sentences \cite{plattner_priming_2018}. 
% Priming as a method was developed by Redelmeimer et al., \cite{redelmeier_memories_2003} in their paper, \textit{‘Memories of colonoscopy: a randomized trial’}. 
% The paper establishes how introducing small interventions in the feeling/experience of pain can change the memory of pain. 
Building on memory recall in psychological sciences \cite{redelmeier_memories_2003}, priming seeks to produce an alterity in a participant's experience and sensemaking by introducing a shift in the process of recall. 
Priming is premised on the idea that a participant's memory is not an exact replay of the complete experience but a recall of  selected moments. 
HCI has adapted priming to support idea generation in designing creativity support tools \cite{lewis_affective_2011}.

%Priming is primarily used in behavioral psychology to discover interesting findings on human judgments and decision making. Priming affects perceptions and behaviors by increasing cognitive accessibility of specific mental content. In the context of design, it is done by using stimuli such as drawings, videos, neutral or hostile sentences etc. \cite{plattner_priming_2018}. Priming as a method was developed by Redelmeimer et al., \cite{redelmeier_memories_2003} in their paper, ‘Memories of colonoscopy: a randomized trial’. The paper shows how introducing small interventions in the feeling/experience of pain can change the memory of pain. It was built on works in psychological sciences on memory recall. How memory is created not by exact running of the total of the experience but created by recalling selected moments. These moments can be predicted. The method has been adopted by HCI and used to support idea generation and to design creativity support tools \cite{lewis_affective_2011}. 

%%%%%%% I am stopping here. Need to come back. 
\textbf{Understanding}: Understanding is a precursor to generative engagement. Orr \cite{orr_talking_2016} looks at technical training of technicians and their diagnosis of machine breakdown. Diagnosing the breakdown requires narrative construction and negotiation between fragments of information from the customer and the condition of the machine. Suchman \cite{suchman_human-machines_2009} problematizes the notion embedded in the `expert help system' for machine use, which embodies a conception of human action shared by designers, behavioral sciences, and common sense that what people say or do is best understood as the reflection of underlying plans. Suchman shows that this confuses \textit{plans} with \textit{situated action}. Plans are not rigid scripts dictating every move; they are vague guides used to explain action while navigating unpredictable situations, designed to accommodate the unforeseeable contingencies of actual situations of action. Understanding, then, engages a person to produce a participant who is in the process of sensemaking.

\textbf{Generating}: In early PD works, generating developed as a mode of engagement in the UTOPIA project \cite{bodker_utopian_1987}. It presents a `tool perspective' where a computing system is designed as a collection of tools for skilled workers to use by building upon their tacit skills as a basis for analysis and design. In order to develop their technical imagination the users have to gain insight into technical possibilities as well. Thus, generating engages a person to produce a participant who becomes familiar with technical affordances in the design process.

The modes of engagement in PD produce a participant from a person in different aspects of processes.  PD produces an \textit{a priori} to the participant (probing); a participant in the process of memory recall (priming); a participant in the process of sensemaking (understanding); and a participant who is familiar with technical affordances (generating). There are \emph{degrees} of participation and representation afforded by the different modes of engagement but never representation of personhood as a whole.

\section{Methodology}

This research is a comparative case study of three synthetic agents deployed in representational processes. 
The three cases were selected to provide variation along two dimensions. First, is the scope of representation where the cases span local (Ana, a community-voice chatbot in Pittsburgh), enterprise (Synthetic Juror, a legal trial-preparation tool), and global scales (Ask Amina and Ask Abdalla, two AI avatars developed by the United Nations University Centre for Policy Research).
Second, populations represented span community members in local policy consultation, jurors in legal deliberation, and refugees and combatant leaders in humanitarian diplomacy.
We restricted ourselves to cases with publicly accessible documentation sufficient for close reading. 
The three cases also vary in technical architecture (Retrieval Augmented Generation (RAG)-based personas in Ana and Amina/Abdalla; a hybrid neuro-symbolic system in Synthetic Juror) and in interface (chatbot, AI avatar, and Slack integration), but variation along these dimensions was not a selection criterion.

Despite the range in scope, relative to other technological systems and infrastructures, synthetic agents in representational processes are an emerging domain, and the cases we examine are pilots, enterprise trials, and prototypes rather than mature deployments.
We see this as the tractable moment for articulating boundaries, before institutional inertia and sunk investments make them harder to draw.

%This research is structured as a case study investigation using three case studies of political and representational synthetic agents in policy, diplomacy, and legal contexts. These three cases provide an insight into three different scales of political and representational application at local (Ana), enterprise (Synthetic Juror), and global (Ask Abdalla/Ask Amina) contexts. These three cases also allow us to investigate RQ1 in diverse domains where political and representational synthetic agents have been used.

\subsection{Analytical Approach}

The analysis proceeded through close reading of the publicly available documentation for each case. 
This included project reports, blog posts, marketing materials, founder writings, and secondary coverage where available. 
For Ask Amina and Ask Abdalla, our primary source was the UNU-CPR working paper \cite{albrecht_does_2025}, supplemented by investigative journalism \cite{gault_made_2025}. 
For Ana, we drew on the New Sun Rising organizational materials and the Community Voice project documentation \cite{horn_new_2025, new_sun_rising_community_2025}.
For Synthetic Juror, we drew on the company's published blog posts and product documentation \cite{doc1_abbey_legal_2024, doc3_abbey_syntheticjuror_2024, doc2_hoogerhuis_beyond_2024, juror_about-_2025}.

Rather than conducting a formal thematic analysis, we worked through iterative paired discussion. 
The first and second authors collected and read the documentation, and developed initial codes for each case. 
The four authors then discussed the cases in pairs, comparing codes across cases and refining them through successive rounds. 
We iterated until our codes were stable across cases, focusing on how each synthetic agent framed its problem context, curated its data, designed its user encounters, and evaluated its outputs.
These four dimensions emerged as analytically useful through the iterative comparison and now structure the Findings section.

While discussing the codes, the limited focus on processes prompted our examination and subsequent application of PD. 
In particular, \citet{sanders_framework_2010} helped us gain a process-oriented vocabulary to set against the outcome-centric evaluations that dominate the existing literature.

\subsection{Positionality}

We are four researchers based in Global North, three of us originally come from the Global South. 
The first and second authors are graduate students working at the intersection of Science and Technology Studies and Human-Computer Interaction (HCI).
The third author is a faculty member with expertise in information science, technology ethics and law who leads action-oriented research projects. 
The fourth author is a faculty member in HCI who has conducted community-based participatory research over the last decade in both the Global South and the Global North. 

Our shared concern in pursuing this work is the over-reach of technology into representational contexts, particularly where the populations being represented are already structurally marginalized. 
We are aware of the tension that we critique tools that are framed as offering support to populations with limited resources, often from our location in well-resourced Global North institutions.
We hold onto this tension. 
Our critique is not that synthetic agents serve these populations poorly (indeed they may speak for these communities well), but that the legitimacy and value of any approach comes from people having the agency and power to speak and act for themselves (i.e., representation). 
We have tried to maintain this distinction throughout the analysis.

We also acknowledge that our materials are drawn predominantly from documentation published by the organizations themselves. 
Much of this material is targeted at potential clients or funders, and some claims may be more aspirational.
We treat the documentation as evidence of how the systems are envisioned and justified.

\section{Case studies}

The three synthetic agents selected for this research are situated in the context of policy, diplomacy, and legal jury. Collectively, they constitute synthetic agents in political and representational contexts at three different scales: local (Ana), enterprise (Synthetic Juror), and global (Ask Abdalla/Ask Amina).

% Amina/Abdalla and Ana Chatbot are variants of LLM-based retrieval-augmented generation (RAG) implementations. They depend on a curated knowledge base for mimicking personas representative of the target population in their respective contexts. Synthetic Juror is a hybrid neuro-symbolic implementation where the human elements of communication are handled by an underlying LLM model while the enforcing of legal standards is handled by the symbolic reasoning layer acting as a logic engine.

% The interface and affordances vary across the three synthetic agents. Amina/Abdalla are AI avatars that can engage in a voice conversation. Ana is designed as a Chatbot interface, and Synthetic Juror gets integrated into the Slack interface of the enterprise client where the synthetic agents interact as individual personas on the Slack channels. 

\subsection{Ask Abdalla/Ask Amina}

Ask Amina and Ask Abdalla are two AI avatars that were created as an experimental tool by the United Nations University Centre for Policy Research (UNU-CPR) for exploring use of AI in Conflict Prevention.
Both are RAG-based implementations with curated knowledge bases targeted at specific populations, presented through voice-conversational avatars.
Amina is positioned as a digital representation of a refugee living in a camp in Chad, and Abdalla as a digital replica of a combatant leader in the Rapid Support Forces (RSF), a group active in the southeastern part of Sudan from which many refugees are fleeing.

The use-case of these avatars is justified by citing the difficulties in recruiting participants for diplomatic consultation and deliberation.
The documentation argues that current methods, including surveys, focus groups, and questionnaires, are time- and resource-intensive, and that participation can be limited by time constraints, geographical remoteness, or unwillingness to participate.
It also notes that respondents may have ``\textit{an incentive to provide false or incomplete information when replying, especially if respondents believe that a certain type of answer may benefit them in some way}'' \cite[p.~4]{albrecht_does_2025}.

The stated goals for the two are operational, which the document states as \textit{``if Amina works, 'her' rapid responses could be of great value. For example, they could be used to quickly make a case to donors (often in very different locations and with very little time) on what population needs to be prioritized when earmarking aid to the region. If Abdalla works, 'his' responses could help negotiators and mediators prepare for more targeted real-world engagement``}\cite[p.~8]{albrecht_does_2025}.

\begin{figure}[H]
    \centering
    \includegraphics[width=0.75\linewidth]{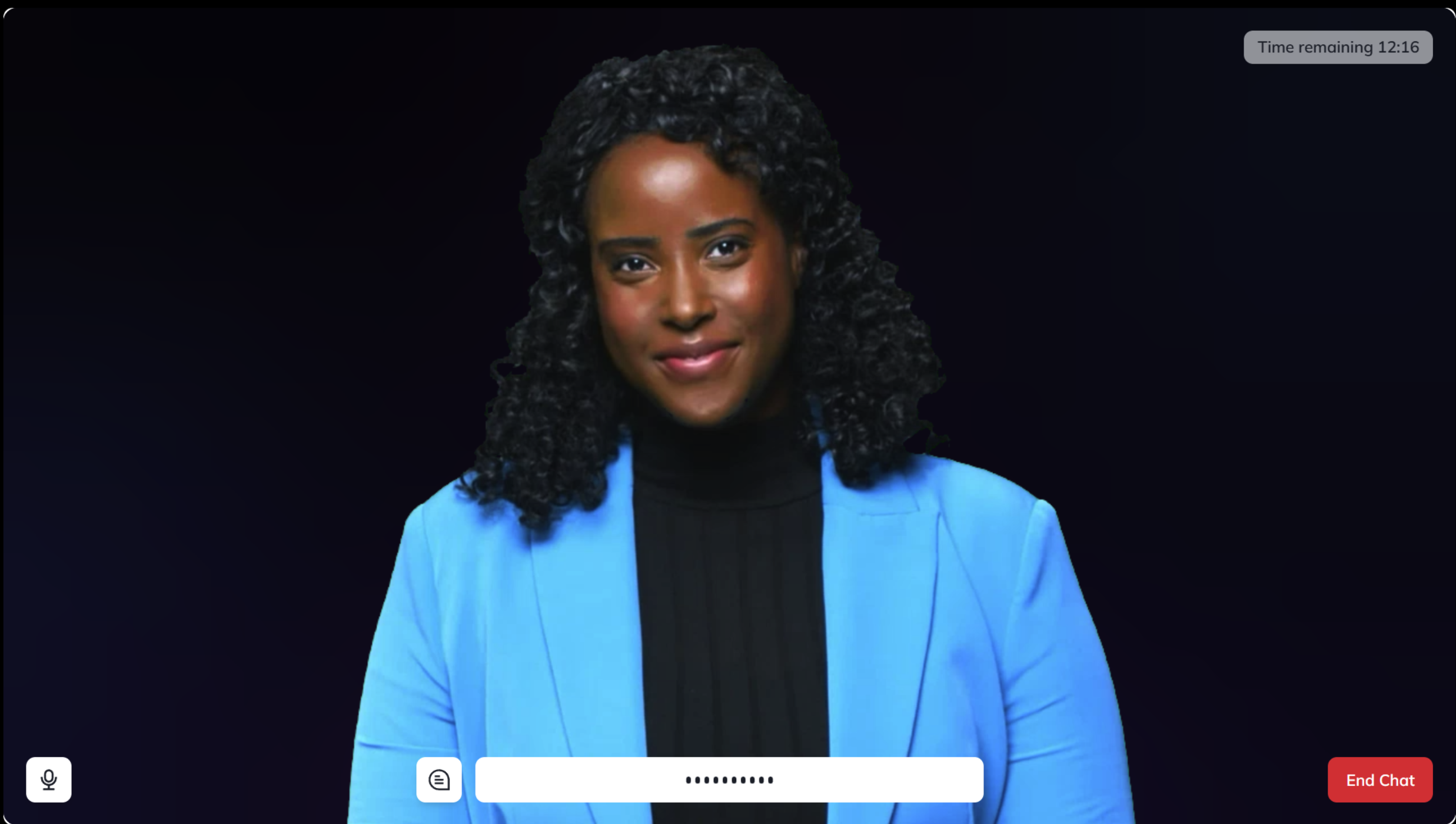}
    \caption{Screenshot of the AI avatar Amina \cite{gault_made_2025}.}
    \label{fig:Amina}
\end{figure}

The development team also expresses an intention to extend the synthetic agents to group deliberation, integrating AI agent-generated personas of stakeholders, including community leaders, government officials, and military leaders, into diplomatic negotiation and mediation simulations \cite[pp. 4]{albrecht_does_2025}.
The two avatars were briefly online for public use but are no longer accessible.

\subsection{Ana (New Sun Rising)}

Ana is a chatbot created by New Sun Rising, a Pittsburgh (Pennsylvania, USA)-based organization, as part of their virtual support center.
Like Amina and Abdalla, Ana is a RAG-based implementation, but rather than mimicking a regional demographic persona, it surfaces quoted excerpts and synthesized insights from a community dataset. 
The underlying dataset is the organization's Community Voice project, which gathers feedback and testimonials from community leaders and members on the present strengths and challenges of their community, as well as their hopes, fears, and expectations for the future \cite{new_sun_rising_community_2025}.

\begin{figure}
    \centering
    \includegraphics[width=0.75\linewidth]{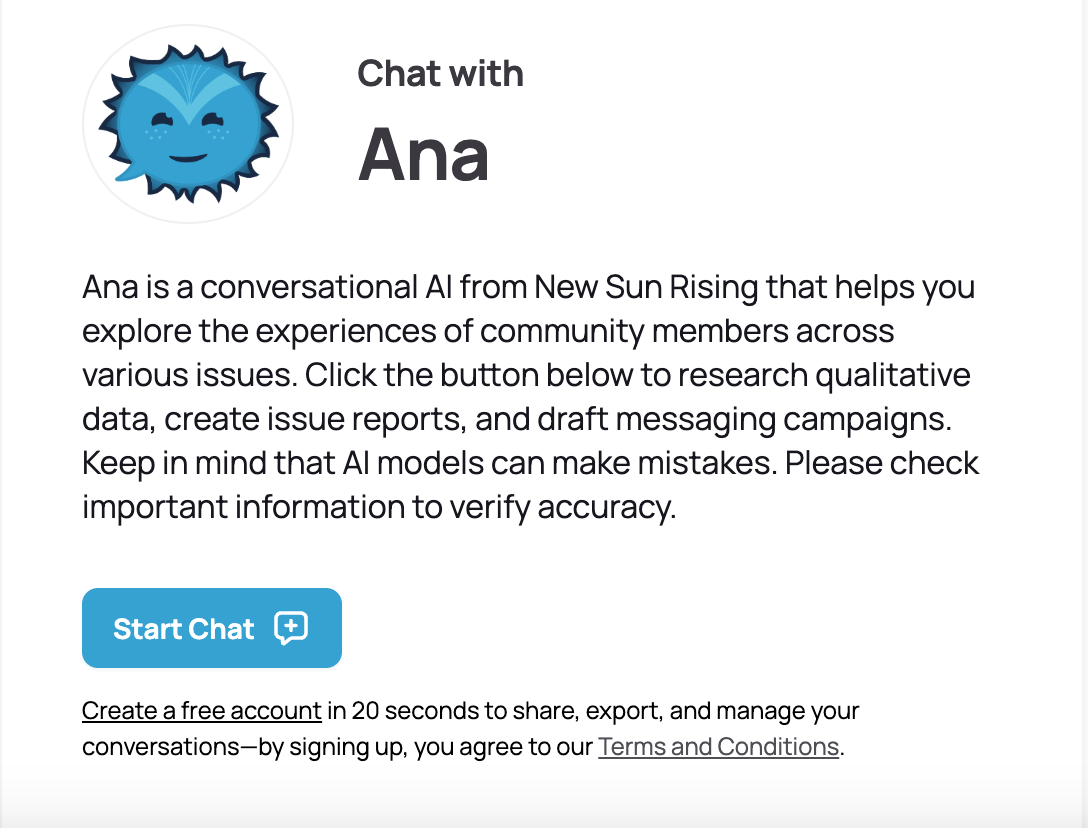}
        \caption{Screenshot of the AI chatbot Ana \cite{new_sun_rising_pittsburgh_ai-driven_2025}.}
        \vspace{-1em}
    \label{fig:Ana}
\end{figure}

The organization coded the dataset using its Impact Framework (mapped to the UN Sustainable Development Goals), a Resource Framework (Community Capitals), and sentiment analysis \cite{new_sun_rising_community_2025}. 
Ana \textit{``...specializes in data analysis and communication, assisting users in exploring the Community Voice qualitative data-set to enhance grant proposals, communications, and advocacy campaigns``} \cite{horn_new_2025}. 
Ana \footnote{Accessible at: \url{https://vibrancy.newsunrising.org/}} is targeted towards non-profits and local policy makers who require inputs from community members in order to make better informed decisions. 
It acts as an interface for exploring and discovering existing data relevant to the query.

\subsection{Synthetic Juror}

Synthetic Juror is a legal technology platform that utilizes a hybrid neuro-symbolic architecture to simulate juror behavior, reasoning, and decision-making. 
The LLM layer handles the ``human'' elements of communication, processing case files and evidence to understand natural language, and patterns in emotional nuances \cite{juror_about-_2025}. 

The system aims to extract deep patterns from the psychographic and demographic data provided by its integrated workbench, `Lairm', to create synthetic juror profiles. 
Legal teams can use these to replace traditional mock trials with simulations that include geo-specific modeling, adjusting to regional attitudes, and narrative testing to identify resonating emotional appeals. 
The company emphasizes the geo-specific dimension as a distinguishing feature, claiming that the platform ``understands the cultural, demographic, and regional nuances that shape juror perspectives'' and lets attorneys tailor  strategies to address biases and emotional triggers specific to the jurisdiction \cite{juror_about-_2025}.

% The symbolic reasoning layer of acts as a logic engine to implement legal standards and jury instructions. \textit{``The platform is built on neuro-symbolic integration, which merges symbolic reasoning with neural networks. This means it mimics how real jurors process evidence, interpret legal standards, and apply personal biases. \cite{juror_about-_2025}``} This provides a structured sequence of reasoning by applying rules of evidence from a legal standpoint. The system facilitates running thousands of simulations. It is integrated with the Slack interface of the client and the synthetic agents interact as personas on the Slack channels. Synthetic Juror is an enterprise software available for trial upon request sent to the company.

The platform supports running thousands of simulations and is integrated into the Slack interface of the enterprise client, where synthetic agents interact as individual personas across Slack channels.
Synthetic Juror is an enterprise software available for trial upon request sent to the company.

\section{Findings}

%% Four steps of taking a representation
%Across the three cases, we trace a four-step process through which synthetic agents become a representational entity. 
%These four steps mirror the stages of AI agent development process. 
% The process begins with how the problem of participation is framed, continues through how data is drawn to create a persona, which then shapes how users are invited to encounter the system , and culminates in how the resulting system is evaluated. 
% We argue that, together, these manufacture personhood. 
Across the three cases we trace a four-step process (framing the problem, curating data, designing the encounter, evaluating validity) through which synthetic agents become a representational entity.

\subsection{Manufacturing the Problem for Synthetic Agents}
\label{sec:problemManufacture}

Establishing the need for a synthetic agent product is the first stage that enables conceptualization of a field of intervention in the representational process. While synthetic agents are generally proposed as a solution to the difficulty in recruiting participants for consultation, each of them speaks to a different problem. 

For Ask Amina and Ask Abdalla, the stated objective is to enable humanitarian and relief workers to ask questions about refugee experiences, needs, and sentiments, and to receive responses that closely mirror real refugee perspectives \cite[p.~3]{albrecht_does_2025}. 
Amina is positioned to enhance the feasibility of needs assessments for UN staff in determining the type of intervention \cite{albrecht_does_2025}, placing the synthetic agent in a role that makes allocative choices about whose interests come first.
Ask Abdalla, on the other hand, is designed to let diplomats, negotiators, and mediators practice their skills with a persona that responds in ways consistent with a real combatant's known behavioural patterns \cite[p.~3]{albrecht_does_2025}.
%\textit{``...creates a digital replica of a combatant leader in the Rapid Support Forces (RSF), a group active in the southeastern part of Sudan from which many refugees are fleeing... The objective of Ask Abdalla is to enable diplomats,  negotiators and mediators to practice their skills with a persona that responds in ways consistent with a real combatant’s known behavioural patterns``} \cite[p.~3]{albrecht_does_2025}.

Collectively, probing and understanding the field is acknowledged as central to the domain enquiry of diplomacy. 
However, the complexities of probing are reduced to a feasibility concern in synthetic agents. 
The framing focuses on manufacturing a ``person-who-probes'' instead of the process of probing. 
These two AI avatars are, in fact, proposed as ``Anthropologist Agents'' that can \textit{``act as `investigators' dedicated to studying a person and their context ... they must become similar to anthropologists''} \cite[pp. 6]{albrecht_does_2025}.
They further list  four skill sets for these agents, covering literature review, fieldwork, culturally responsive analysis, and continuous knowledge base updates \cite[p.~6]{albrecht_does_2025}.
This frames the synthetic agent as an expert on the lived experiences of the community whom it claims to represent, and importantly, generates a tension between the two incompatible positionalities, as an embodied refugee or combatant on the one hand, and the external observer and investigator on the other. 

%The documentation on Amina and Abdalla also lists the underlying goals of an anthropologist agent \cite{albrecht_does_2025}:
% ``\textit{``...anthropologist agents – just like real anthropologists must have four key skills: 1) They must know how to autonomously connect to the right existing literature and past research conducted on the population... 2) They must know how to collect and categorize all relevant cultural artifacts produced by the people...conduct fieldwork. 3) They must be able to critically analyse all this information and then organize it in a way that is culturally responsive...4) They must be able to continuously update and reorganize this knowledge base, based on the evolving information environment...``}\cite[p.~6]{albrecht_does_2025}.``

Ask Amina/Abdalla seeks to configure a mechanized probe tool which can be attributed the legitimacy of an anthropologist expert. 
%In contrast, Ana \textit{``...specializes in data analysis and communication, assisting users in exploring the Community Voice qualitative data-set to enhance grant proposals, communications, and advocacy campaigns``} \cite{horn_new_2025}. 
In contrast, Ana draws out synthesized insights and quoted interview excerpts from the Community Voice dataset \cite{horn_new_2025}. 
In doing so, it makes moral choices on behalf of its users by prioritizing some voices over others on grounds of relevance, even as it does not claim to embody a persona.
%The Community Voices qualitative dataset contains interviews and surveys of community leaders and community members.
% Ana is targeted towards non-profits and policy makers who require inputs from community members. It acts as an interface for exploring and discovering existing data relevant to the query. Generating synthesized insights and quoting interview excerpts in its generated responses makes moral choices available to Ana by prioritizing some voices over others on the measures of relevance.

Synthetic Juror intervenes in legal trial preparation for lawyers as an enterprise solution addressing the feasibility challenges of conducting a mock trial. 
The company frames traditional trial preparation as relying on mock trials, focus groups, and intuition, and positions itself as bringing \textit{``a new level of precision by simulating thousands of jurors' reactions using AI.''} \cite{doc1_abbey_legal_2024}.
The system claims to model \textit{``how real jurors think, incorporating their biases, emotional triggers, and regional quirks ... These aren't just static profiles. They're active participants ...You can present any evidence to them  (depositions, expert witness, potential outcomes, interrogatories, motions, arguments and the list goes on) ... they respond ... with nuanced feedback that mimics how real jurors think``} \cite{doc1_abbey_legal_2024}.
%Integrated into Slack, this tool allows lawyers to present evidence, test arguments, and receive real-time feedback from AI-generated jurors, tailored to specific jurisdictions.'' \cite{doc1_abbey_legal_2024}.
The appeal to legitimacy is the ability to narrativize and imitate the case context like a person sitting in the jury.
This claim invites interrogation: can synthetic jurors be probed for bias and impartiality, which is the established procedure of \textit{voir dire} in legal jury deliberations?

% SyntheticJuror is focused on simulating life-like reactions calibrated to a selected jury demography that matches the expected jury most closely. \textit{``...SyntheticJuror simulates juror reactions using AI... It doesn't just spit out predictions. It models how real jurors think, incorporating their biases, emotional triggers, and regional quirks... These aren't just static profiles. They're active participants...You can present any evidence to them  (depositions, expert witness, potential outcomes, interrogatories, motions, arguments and the list goes on)... they respond...with nuanced feedback that mimics how real jurors think``}\cite{doc1_abbey_legal_2024}. The appeal to legitimacy of synthetic agents in mock jury deliberation is the ability to narrativize and imitate the case context like a human person sitting in the jury. The claim to legitimacy can be interrogated: can synthetic jurors be probed for bias and impartiality; which is an established procedure in legal jury deliberations as \textit{voir dire}?

Across the cases, synthetic agents address an identified feasibility gap by targeting needs like quick surveys, data-driven insights, and variety of mock-simulation of personas. 
While the promise of quick, real-time qualitative insights are the core offering, it could have been delivered as reports or dashboards. 
The personas appear as a separate layer that anthropomorphizes the interaction through narrativization.

% (Discussion points connecting to this subsection:)
%  effort at - Appearing legitimate– anthropologist agent - ability to mimic real human jury-like emotional responses and triggers
% Identity flattening, Misportrayal (portaryal from out-group than in-group) and identity essentialization 
%  The only way of measuring is Believability and plausability 

\subsection{Manufacturing Persona through Data Access}
\label{sec:personaManufacture}

The synthetic agents across representational contexts use qualitative datasets, surveys, and reports in their RAG implementation. 
%The documentation on Amina and Abdalla says that the dataset contains \textit{``...existing literature and past research automatically collected and organized by the anthropologist agent for this pilot study includes several dozens of surveys, reports, articles, issue briefs and other studies conducted among or about refugees and combatants in Chad/Sudan by various UN agencies, multilateral organizations, think tanks, research organizations and NGOs. These data sets are separate for Amina and Abdalla’’}\cite[p.~8]{albrecht_does_2025}.
For Amina and Abdalla, the knowledge base contains surveys, reports, articles, and issue briefs from UN agencies, multilateral organizations, think tanks, and NGOs working with refugees and combatants in Chad and Sudan \cite[p.~8]{albrecht_does_2025}.
A separate persona-curation dataset draws from cultural artifacts of the community being mimicked, including  \textit{``... seminal works of literature, folkloric traditions and key religious texts, but also items from the culture's  contemporary digital footprint, including local news media, social media activity and online diaspora community discussion forums.''} \cite[p.~8]{albrecht_does_2025}. 
The voice-avatar also attempts to mimic the tone and filler-sounds between sentence pauses. An investigative article found that the contemporary digital footprint likely included voice and video recordings of community members \cite{gault_made_2025}.
% A short voice conversation with Amina was available in an investigative article by 404media \cite{gault_made_2025}. It indicates the possibility that the ``contemporary digital footprint`` mentioned in the dataset also includes the voice and video recordings to train the voice-avatar.

Persona curation is also central to Synthetic Juror, given its intention to simulate a diverse mix of jury positionalities. As the document states, it \textit{``uses advanced patent-pending technology to simulate juror responses based on a broad range of data''} \cite{doc2_hoogerhuis_beyond_2024}.
%creating highly realistic and diverse synthetic juror profiles... to model juror behavior across different demographics, psychological profiles, and legal scenarios``}\cite{doc2_hoogerhuis_beyond_2024}. 
Another document indicates \textit{``the synthetic character samples are built using a combination of public data, subscription databases, your firm's proprietary information, and client-specific data, ensuring a high degree of realism and relevance''} \cite{doc3_abbey_syntheticjuror_2024}.

Both Amina/Abdalla and Synthetic Juror manufacture personas through a specifically curated dataset that is separate from their knowledge base. 
Ana, in contrast, draws its data from the Community Voice project curated by the same organization to \textit{``gather feedback and testimonials both digitally and in person''}, in which people were \textit{``invited to share their perspectives on the present strengths and challenges of their community, as well as their hopes, fears, and expectations for the future''}  \cite{new_sun_rising_community_2025}.
The community data was coded by the organization before using it to train the chatbot model \cite{new_sun_rising_community_2025}. 
Unlike Amina/Abdalla and Synthetic Juror, the chatbot Ana does not appear to mimic a region-specific demographic persona.

This data work bypasses the participatory mode of understanding, where sensemaking is constructed through narrative negotiation between participant, researcher, and context \cite{suchman_human-machines_2009, orr_talking_2016}.
Synthetic agents collapse sensemaking into static dataset curation. 
Datasets drawn from institutional sources reflect the plans and blueprints of the institutions that collected them, leading to an over-representation of institutionally prioritized data. 
Paradoxically, the embedded presence of institutional blueprint sabotages the very grounds that necessitate ethnographic inquiry for understanding to complement and evaluate the institutional paradigms.
There is also a clear demarcation between the datasets that generate insights, and the datasets for persona-curation which focuses on drawing positionality from demographic data and the ability to mimic personhood in the context of the prompt.

% [Discussion Points for this subsection]
%The first epistemic risk is category error. Because it mistakes a synthetic predictive next-word text generator for an analytical aid (Nguyen & Welch, 2026).
%Fifth epistemic risk is titled as, ‘Oracle Effect’. Machine-generated outputs are often mistakenly viewed as neutral and unbiased due to their technological origin, thus, as more reliable than human analysts prone to error. This misplaced trust creates a false sense of analytical depth, masking the models’ limitations (Nguyen \& Welch, 2026).

% Rin Su's paper identifies a few problems with the model. especially the tail cases. Another paper that talks about qualitative research and why can't be accept Chatbot's advice in that domain as well. Chatbot can never say things only based on data provided. they mix data with other data. 

\subsection{Manufacturing Encounters for User Interaction}
\label{sec:encountersManufacture}

The three cases offer widely varying affordances to target users.
The documentation on Amina and Abdalla justifies the AI avatar persona by contrasting it with NLP-based sentiment analysis, where NLP can only examine existing opinions, personas can engage in dialogue about hypothetical scenarios \textit{``offering predictive insights into how people might respond to future activities and actions''} \cite[p. 4]{albrecht_does_2025}.  %\textit{``By enabling dynamic conversations rather than just passive data mining, personas provide a considerable advantage over commercial AI-based online sentiment analytics platforms that rely on natural language processing (NLP). While NLP can only examine existing opinions, these personas can engage in rich dialogue about hypothetical scenarios, offering predictive insights into how people might respond to future activities and actions``}\cite[p.~4]{albrecht_does_2025}. The comparison with NLP-based sentiment analysis establishes a clear distinction between analysis of existing data, and predicted/hypothetical data; emphasizing that synthetic agents manufacture interaction surfaces with predicted scenarios. While probing and priming could potentially enable the revealing of asymmetric power relations, such modes of enquiries are no longer viable for hypothetical and predicted scenarios.
This establishes a clear distinction between analysis of existing data and predicted or hypothetical data, and importantly, it positions synthetic agents as interaction surfaces for predicted scenarios.
While probing and priming could potentially reveal asymmetric power relations, such modes of inquiry are no longer viable for hypothetical and predicted scenarios. Priming is thereby bypassed since there is no recall of lived experience to shift.

%SLACK INTERFACE - if space permits
% \begin{figure}[H]
%     \centering
%     \includegraphics[width=0.99\linewidth]{Paper_Content/Synthetic Juror_AI.png}
%     \caption{Screenshot of the Slack interface of Synthetic Juror \cite{doc1_abbey_legal_2024}.}
%     \label{fig:SyntheticJuror}
% \end{figure}

Synthetic Juror extends this affordance to legal practitioners through integration with Slack. \textit{''All of this happens right in Slack. Lawyers can @mention a simulated juror just like they would a colleague. They can create sub-channels for different aspects of the case... use Slack's huddle feature to do mock cross-examinations with AI witnesses...''} \cite{doc1_abbey_legal_2024}.
It further claims that the \textit{``Lairm Workbench includes a sophisticated character creation tool that allows you to build incredibly detailed synthetic individuals for your case simulations''}\cite{doc3_abbey_syntheticjuror_2024}.

This affordance directly bypasses priming as it operates in actual jury deliberation.
Priming in juries takes the form of implicit bias instructions designed to make jurors reflect on their own biases, and the \textit{Sommers effect} \cite{sommers_racial_2006} shows that racial diversity among jurors primes individuals to exchange a wider range of information. 
By offering the ability to calibrate specific demographic positionalities at will, Synthetic Juror removes the moment of reflexivity that is integral to priming and replaces it with essentialized and flattened identities.

% ``\textit{The Lairm Workbench includes a sophisticated character creation tool that allows you to build incredibly detailed synthetic individuals for your case simulations here are some sample characters:\\
% Jurors with specific demographic, psychographic, and attitudinal profiles...Witnesses with varied backgrounds, personalities, and credibility factors...Judges and arbitrators with detailed judicial philosophies and decision-making patterns...Opposing counsel with specific litigation styles and strategic approaches \cite{doc3_abbey_syntheticjuror_2024}}``

Ana's interaction is a chat interface. 
That leaves a wide-open possibility for interaction and risks confusing users. 
The organization recognized this risk and in response, they delivered an information session that included sample prompt templates \cite{new_sun_rising_pittsburgh_ai-driven_2025}. 
A representative template asks Ana to identify the primary themes in the Community Voice data relevant to a specified organization, issue, and audience, and to draft a 500-word summary including quotes from community members \cite{new_sun_rising_pittsburgh_ai-driven_2025}. 
Since templates scaffold the user toward indexing and summarization instead of encounter with a person, there is no moment of recall to prime.

% ``\textit{Draft a Community-informed Issue Summary (with ANA)}
% \textit{My organization [ORG NAME] is focused on improving the [GOAL/MISSION] of [COMMUNITY NAME] residents who are negatively impacted by [ISSUE]. What are the primary themes that the Community Voice data includes from residents' experiences that are related to my work?...}\\
%     % \item \textit{What are the top three Sustainable Development Goals represented in the Community Voice data about these themes [or SPECIFY ISSUE] that relate to [ORG NAME]?}
%     % \item \textit{What are the themes around the Community Capitals represented in Community Voice data about these issues [or SPECIFY ISSUE] that relate to [ORG NAME]?}
% \textit{Provide specific quotes from residents supporting the most common themes about these issues [or SPECIFY ISSUE] from the Community Voice data set.}\\
% \textit{Use this information to draft a 500 word summary of the community voice data on [ISSUE] for [AUDIENCE], including quotes from community members and the most compelling data that would convince them to get involved and take action \cite{new_sun_rising_pittsburgh_ai-driven_2025}.}``

We see these design affordances as different approaches of engaging the user in speculative inquiry and visualization. 
The synthetic agents either lead the user into visualizing an actual embodied individual (e.g., a refugee from Chad), the opinions of an individual (e.g., Slack messages of a jury member), or the reported insight of an individual (e.g., an insight reported by Ana). 
In all cases, the synthetic agents give an impression of personhood while mechanically drawing data and inferences from their existing knowledge base.

\subsection{Manufacturing Validity with Technical Measures}
\label{sec:validityManufacture}

We note that the validity of the synthetic agents is justified by manufacturing the problem context. 
However, a parallel move occurs at the level of outcome evaluation. 

Synthetic Juror claims \textit{``99\% accuracy rate in predicting case outcomes. That's not just a marketing claim. It's based on comparing the system's predictions to actual trial results''} \cite{doc1_abbey_legal_2024}. 
The documentation on Amina describes a study where Amina was fed 20 questions drawn from four surveys not included in her knowledge base, with the delta between her responses and the actual human responses measured as the indicator of accuracy; \textit{``she correctly answered 16 out of 20 questions, achieving an 80 per cent accuracy rate'' \cite[p.~7]{albrecht_does_2025}}.
In contrast, \textit{``Abdalla’s responses, [were] qualitatively assessed. A brief conversation between the two [will] also be qualitatively assessed}'' \cite[p.~7]{albrecht_does_2025}.

%It signal a trial study where the predicted outcomes by the simulated synthetic agent jury were measured against the jury decisions in an actual trial with the jury demographics matching those of synthetic agents.

% Documentation on Amina and Abdalla mentioned \textit{``Amina, will be fed questions that were asked to the same population in real life – the answers to which are not included in the knowledge base – and the delta between the avatar and the actual human responses will be measured. The smaller the delta, the more accurate the avatar... This study evaluated Amina’s representativeness using 20 questions drawn from four distinct surveys, none of which were included in her knowledge base: the SENS Nutritional Survey (4 questions), Post-distribution Monitoring Report of Food Assistance in Refugee Camps (3 questions), Norwegian Refugee Council’s “War in Sudan” (8 questions) and UNHCR’s “Sudanese Emergency” (5 questions). Analysis of Amina’s responses revealed that she correctly answered 16 out of 20 questions, achieving an 80 per cent accuracy rate``}\cite[p.~7]{albrecht_does_2025}.

The available documentation does not describe any evaluation method for Ana.
They measure the similarity of synthetic outputs to human outputs (a delta, a percentage accuracy, a qualitative comparison) and treat that similarity as the warrant for representational validity. 
This evaluation logic bypasses the participatory mode of generating, where the process of setting up a deliberative mechanism is as integral to legitimacy as the artifact it produces. 
PD tools for generating design through group deliberation reveal that outcomes are not automatic products but are closely tied to the underlying process \cite{simonsen_infrastructuring_2020}. 
Outcome-centric evaluation makes the process invisible, and importantly  it makes the process un-evaluable.

\section{Discussion}

\begin{figure*}[ht]
    \centering
    \includegraphics[width=1\linewidth]{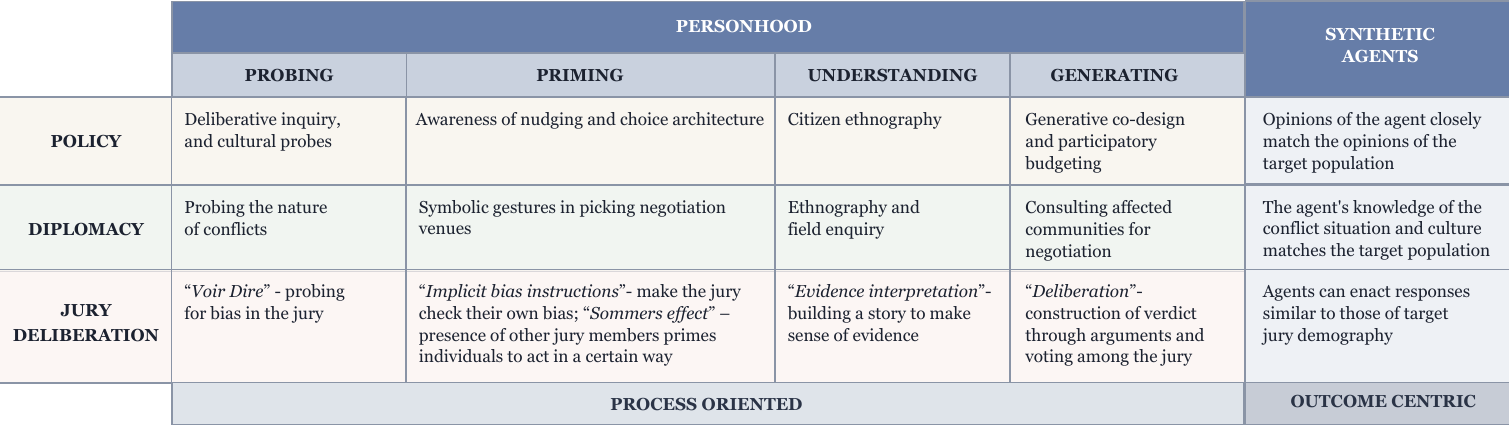}
    \caption{Synthetic agents seek to replace a complex representational process.}
    \label{fig:Participation}
\end{figure*}

The synthetic agents in our three cases are early-stage and experimental, and they reflect the broader nature of representational synthetic agents as an emerging domain.
This is the moment when boundaries are most tractable, before institutional inertia and sunk investments make them harder to articulate. 
In this section, we make explicit what is at stake in the four-step manufacture we traced in Findings, and articulate what synthetic agents categorically cannot do.

% Extension of synthetic agents to representational processes presents a complex problem. While the cases picked in this study are early stage explorations and experimental in nature, they reflect the nature of representational synthetic agents as an emerging domain. In that sense, our early-stage study seeks to establish directions, concerns, and boundaries for such explorations.

% The AI ethics community has previously witnessed algorithmic and predictive technologies being introduced as experimental implementations and then rapidly scaled into application. A well-known case being the COMPAS (Correctional Offender Management Profiling for Alternative Sanctions) predictive decision-support system implemented in the judicial courts on the basis of internally conducted outcome-centric evaluations \cite{larson_how_2016}. \fixme{More literature on jury systems to be added here if space permits}.

% Establishing early boundaries for such precedents is important before institutional investments start determining the momentum of the field.

\subsection{Bypassing Representational Process}

As the Findings showed, the need for synthetic agents is manufactured by emphasizing the difficulty of recruiting participants as the central bottleneck of representational processes, for which a sophisticated mechanical process is offered as the solution. 
This amounts to a ``machinal bypass'' \cite{kaplan_machinal_2025} of a multidimensional process where representation is integral to any claims of legitimacy for an institution.

The central function of representation in participatory processes is the fidelity to the participant population \cite{premananda_non-venatus_2026}. 
From this view, the limited access to participants is not a mere problem to solve with a single solution. 
It is a symptom of deeper structural failures of representation, and  PD scholarship has long questioned whether slow prototyping or feasibility are bottlenecks at all \cite{elavsky_genai_2026}.

Existing evaluation frameworks for synthetic agents (\textit{believability} \cite{xiao_how_2024}, \textit{plausibility} \cite{agarwal_faithfulness_2024}, \textit{faithfulness} \cite{maden_three_2026}, \textit{algorithmic fidelity} \cite{argyle_out_2023}) all measure similarity between human and synthetic responses, treating outcome resemblance as the warrant for representational validity.
The table in Fig. \ref{fig:Participation} contrasts this with PD modes of engagement applied to the same three domains.
PD treats representation as a process whereas synthetic agents reduce it to an outcome.
Let us consider the strongest version of the case these frameworks make, a synthetic agent whose responses are statistically indistinguishable from those of the population it stands in for. 
Even at this limit of perfect fidelity, the agent delivers the resemblance of participation and not participation itself, and, as established above, it is the participation that legitimacy requires.
The bypass thus becomes structural since existing outcome-centric evaluation makes emphasis and orientation on the processes impossible.

\subsubsection{Reckoning Replaces Judgment}

%% Aditya, this is too much background for a discussion
% It is helpful to borrow the distinction between ‘reckoning’ and ‘judgement’ \cite{weizenbaum_computer_1976, smith_promise_2019} to explore the underlying fallacy of using synthetic agents in representational processes. Building on Weizenbaum's (1976) work Smith (2019) distinguishes between two forms of reasoning namely, reckoning and judgement where reckoning is the act of estimating or a sequence of calculations which can be sophisticated but can still be broken down in terms of simple logical rules within a well-defined space. This form of intelligence is used for playing games like chess. On the other hand, judgment is a completely different form of intelligence that requires dispassionate and deliberate thought guided by ethical commitments and awareness of consequences resulting in actions appropriate to an undefined situation. In our case study context, jury deliberations constitute an essential process of 'evidence interpretation' (Fig. \ref{fig:Participation}), where the jury collectively deliberates and constructs a story from evidence in order to arrive at a verdict. The evidence interpretation cannot be reduced to the verdict itself.

The distinction between \textit{reckoning} and \textit{judgment} clarifies why this bypass is consequential \cite{weizenbaum_computer_1976, smith_promise_2019}.
Reckoning is estimation or calculation that can be sophisticated but operates within a well-defined space governed by simple logical rules. 
For instance, a game of chess is reckoning. 
Judgment is a different form of intelligence requiring dispassionate and deliberate thought guided by ethical commitments and awareness of consequences, producing actions appropriate to an undefined situation \cite{weizenbaum_computer_1976}.
In our case studies, jury deliberation is paradigmatically a process of judgment. 
The jury constructs a story from evidence to arrive at a verdict, and importantly, the evidence interpretation cannot be reduced to the verdict itself.

LLMs are reckoning machines due to their epistemic method of reproducing the cosine distances present in their vector-embeddings in the high-dimensional space. 
These vector-embeddings are abstracted from the past training data and hence are constrained by them. Related literature on simulation of judgment in LLMs argues that the very notion of judgment is operationalized when decisions are made by LLMs, where similarity of outputs `appear as alignment` but may conceal deeper epistemic shift where normative reasoning is replaced by statistical approximation. The authors ask what heuristics are encoded in this process of delegation \cite{loru_simulation_2025}? 
Such a system cannot orient itself towards the concerns of the future or the consequences of its actions.
As we saw in the Findings section, synthetic agents mimic human personhood by imitating human responses and even have a separate dataset curated just for the purpose of imitation. 
The deliberate use of anthropomorphic terminology, such as  ‘reasoning’, ‘thinking’, `hallucinations', indicate an attempt to conflate the human and machine processes.
Importantly, it stretches the logic of context representation to claim that reckoning machines are capable of judgment. 

This has deep consequences beyond poorly made moral choice. 
Allocative choices, such as whose needs to prioritize when earmarking aid (Amina/Abdalla), and choices about which voices to prioritize (Ana), are political judgments. %made without any reciprocal relationship of accountability, or the ability to experience consequences. 
In fact, when they are represented as output of a calculation engine, they are often removed from the domain of contestable political judgment and relabeled as technical fact. 
It launders judgment, conferring on the synthetic agent the functional status of a moral actor while remaining categorically incapable of intentionality or of entering into reciprocal relations as a subject of rights and duties.
This relocation of political choice into the seemingly neutral output of a machine is what later allows those in authority to defer accountability for it, a deferral we take up in the next section. 
Substituting a synthetic agent for a human participant is, in the end, the application of a reckoning mechanism in a context that categorically requires judgment.

%Synthetic agents read human action as a language and develop its simplified and stylized representation \cite{stark_animation_2024}. Synthetic agents conflate reckoning with judgment and hence stretch the logic of context representation to claim legitimate agents capable of judgment. It is critical to understand that using synthetic agents as proxy for human participation is equivalent to applying a reckoning mechanism in contexts that requires human judgment.

\subsubsection{Reflexivity Cannot Be Mechanized}

A related bypass occurs at the level of reflexivity.
Synthetic agents often draw their imitation data from ``seed-interviews'' \cite[p.~6]{narasimhan_flight_2026} which collect opinions and associate them with their demographic data. 
In our case studies, Ana draws on the Community Voice project, and Amina/Abdalla draw on social media data and surveys. 
Beyond identity flattening and essentialization \cite{su_toward_2026, wang_large_2025}, this epistemic method amounts to a claim of mechanically reproducing reflexivity.

Participation in representational processes requires reflexivity. 
%It takes away the reflexivity available to a participant in representational processes where they reflect on their positionality when probed, before generating a response. 
When a citizen responds to a policy consultation or a juror weighs evidence, they reflect on their positionality (as a citizen, as a juror) before generating a response. 
This response is shaped in the act of reflection. 
Synthetic agents bypass this moment and replace it with a probabilistic algorithm of mechanical context estimation. 
The extreme case in our findings was the proposed \textit{anthropologist agent} \cite[p.~6]{albrecht_does_2025}, which conflates the positionality of the embodied refugee or combatant with that of the external observer studying them.

Synthetic Juror represents the inverse extreme. 
The platform allows users to curate demographic identities for synthetic jurors and claims the resulting agents respond as people from those positionalities would. 
An individual is reduced to a tuple of social and systemic categories. The training data of foundational models and the RAG implementations built on them are constrained by the computational requirement to converge on the most probable output, which means generated responses identify with the median of the training data and exclude marginalized voices that exist in the tails \cite{su_toward_2026}.
Essentialization is, thus,  structurally produced by these epistemic methods.

Personhood is constituted relationally through political, legal, and moral dimensions that together determine the relationships a person holds with other persons and institutions. Participation, representation, and legitimacy are entangled in this relational notion.
But synthetic agent reproduces the appearance of participation while removing its substance, the process, the judgment, and the reflexivity that make participation legitimate. That is, it ``\emph{manufactures personhood}''. 
When synthetic agents manufacture personhood, they reduce these relational dimensions to essentialized identities and produce an appearance of legitimate participation without the underlying process. No\emph{body} participates.

\begin{figure*}[h]
    \centering
    \includegraphics[width=1\linewidth]{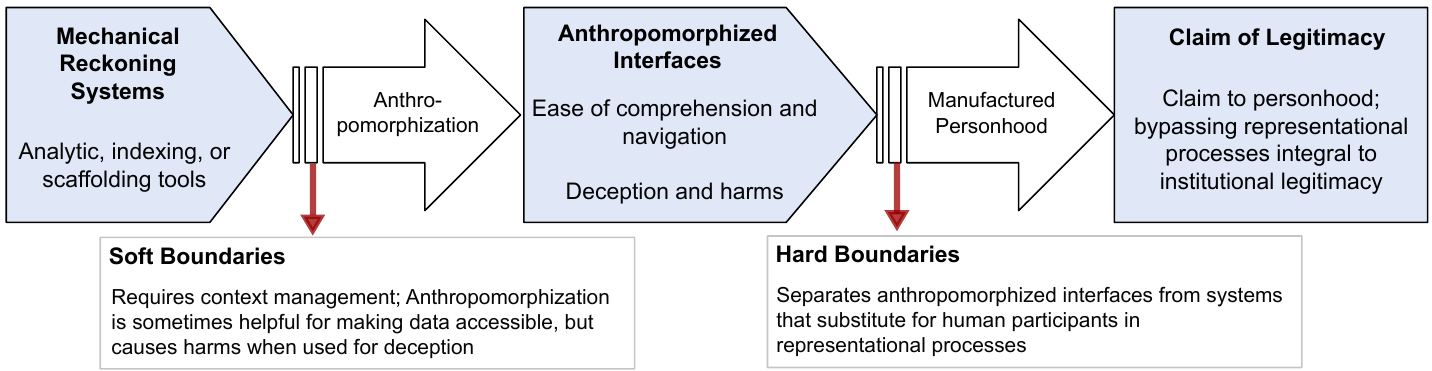}
    \caption{Soft boundaries and hard boundaries to develop oversight for synthetic agents.}
    \label{fig:Boundaries}
\end{figure*}

\section{Boundaries for Design}

The manufacture of personhood enables a further deferral.
When LLMs substitute for human participants, those in authority can claim that representational choices were algorithmically determined by a calculation engine that cannot itself be evaluated. 
This problem has been examined in the context of e-commerce as the principal-agent-third-party problem of trust and verification \cite{riedl_ai_2025}, but it has not been conceptualized for representational decision-making. 
Related work has named the broader pattern an \textit{accountability shield} \cite{timaite_agents_2025} and \textit{algorithmic governmentality} \cite{rouvroy_algorithmic_2013, nayak_boredom_2021}. 
Representational contexts intensify the stakes because the choices are political. 
Drawing boundaries becomes critical to ensure we can center people's power and agency in the political processes.

Drawing on the manufactured personhood we traced above, we propose two boundaries for researchers and designers working with LLM-based systems in representational contexts. 
The first is ``soft'' and context-dependent. 
The second is ``hard'' and categorical. 
We build this distinction because hedged versions of the second boundary erode under pressure from precisely the actors most likely to scale or transfer these systems beyond their experimental scope \cite{larson_how_2016}.

The soft boundary separates LLMs functioning as analytic, indexing, or scaffolding tools from LLMs that present themselves as person-like interlocutors. 
This boundary is contextual. 
Anthropomorphization is sometimes useful. 
For example, it can make complex data more navigable and thus accessible \cite{tan_exploring_2025}. Other times, it is harmful when it naturalizes claims a system cannot make \cite{lifshitz_bc_2024}.

The hard boundary separates anthropomorphized interfaces, which are mere ``evocative animations'' \cite{stark_animation_2024}, from systems that substitute for human participants in representational processes. 
This boundary is categorical and absolute in the sense that no degree of fidelity, believability, or algorithmic sophistication moves a system across it.

We draw this categorical boundary based on the fact that representational processes (e.g., policy consultation, jury deliberation, diplomatic negotiation) derive their legitimacy not from arriving at the right outcome but from being constituted by the participation of persons. The participation is in the service of legitimacy. 
For example, a jury verdict in the US is legitimate because twelve persons from society deliberated and not because a tribunal predicted what they would have said. 
Similarly, a policy consultation rests its legitimacy on having citizens' input. 
% In some cases, it may not be true (e.g., China) but that does not mean it ought not be the case.  
%% So then does this mean we can consult AI outcomes to “correct” our process outcome(s)? No, becuase if we did, then it would dismiss the process itself.

\subsubsection{On the Hardness of Hard Boundaries}
Synthetic agents categorically lack the properties that participation requires, which are consciousness, intentionality, embodied experience, the capacity to bear consequences, and the capacity for reciprocal moral relations. 
These are conditions of personhood that constitute representational processes.

We understand that this may rule out reasonable middle-ground positions that are argued in the literature and popular media. 
\emph{Community-consented synthetic representation} assumes a community can authorize a synthetic agent to be its representation in deliberative settings. 
However, participation is non-transferable. 
No one can consent on another's behalf to having that other's participation substituted, because the participation is the thing legitimacy requires \cite{agnew_workshop_2026}.
\emph{Augmentation-as-supplement} assumes a synthetic agent can speak alongside human participants without displacing them. But the synthetic voice in a representational process is doing representational work, howsoever it may be framed.
\emph{Training simulations for real deployment} assume practice on a simulation transfers to competence in the real thing. 
But what is practiced is engagement with a text-prediction system, and importantly, not engagement with a refugee, a juror, or a combatant leader and all the contingencies and uncertainties that are at play in those contexts.

From our perspective, the single remaining use case is education or training that is explicitly scaffolded as non-transferable, where the discontinuity between the simulated encounter and the real one can be a lesson. 
A diplomat-in-training can encounter a system like Ask Abdalla provided the encounter is framed as engagement with a text-prediction system, and importantly, not as preparation for engaging with actual combatant leaders.
A law student can practice argumentation against a synthetic juror to improve the students' argument skills and not as a way to learn how to engage with the judicial system. 
In these cases, a carefully created scaffold to draw boundaries is necessary.

\subsubsection{Soft Boundaries Require Maintenance}

Of our three cases, Ana sits closest to the soft boundary because it surfaces existing community voices through quoted excerpts.
However, the representational problem remains. 
The chatbot persona, by lending mediated authority to the indexed material, performs legitimation work. 
This instability is intrinsic to soft boundaries.
As community members continue to engage with such a system, as organizational staff cite its outputs in communications, and as it becomes the default interface to the community dataset, the system accrues representational status that was never formally claimed for it.

Indeed, the legal recognition of this drift has been established.
In \textit{Moffatt v. Air Canada}, the British Columbia Civil Resolution Tribunal held the airline responsible for its chatbot's statements, rejecting the argument that the chatbot was a separate legal entity from the company it served \cite{lifshitz_bc_2024}.
The tribunal recognized that the deployment of a chatbot in a customer-facing role makes the chatbot a functional representative.

Maintaining the soft boundary, therefore, requires active design and governance work, and importantly, it is not a property the system possesses on its own. 
Interfaces should clarify and explicitly emphasize the system's role in indexing data rather than being a voice. 
Outputs that quote or summarize community data should route attribution back to the community sources rather than aggregate into a single chatbot persona.
Importantly, evaluation processes should test for representational drift over time and, building on participatory design tradition, ask process-oriented questions. 
This includes asking questions like \emph{Whose voices and processes are informing the system?} A system that is informed by the researcher's analytic process is differently situated than one that purports to involve a community's participatory process. \emph{Does the system create or reduce opportunities for different stakeholders to participate in their own terms?} A system that points researchers toward community voices they had not heard expands participation. 
\emph{What would the represented community say about the system, in conditions where their assessment carries weight?} 
This shifts the evaluative locus from the researcher's judgment of fidelity to the community's judgment of the system's relation to them.
%It is critical to note that none of these questions can be answered at deployment. 

% PD as a trap— Premise underlying participation can never be satisfied by synthetic agents
% We are talking about a tool where the process is mostly pbfuscated in the process (can be a dicsussion point)- outcome oriented vs [rocess oriented
\section{Limitations}

To conclude, given the experimental nature of the cases available for studying, we acknowledge that our case materials are drawn from the documentation published by the organizations. Since a lot of this material is targeted towards their potential clients, some of their claims may be more aspirational than material. The inability to interact with the three AI agents directly is another limitation which could have added more analytical depth to our analysis. Ana chatbot was available for interaction, but its terms of use limit us from discussing the generated responses.

\section{Acknowledgments}
An earlier version of this work was presented at the symposium \textit{'Chat Token Vector: Questioning Models of Language and Neo-Structuralism in AI'} organized by the ERC project AIMODELS at the Department of Philosophy and Cultural Heritage of Ca' Foscari University of Venice in partnership with Cambridge Language Sciences and Cambridge Digital Humanities. We are thankful to the symposium for their valuable feedback. We would like to express our gratitude to Takuya Maeda for the productive discussions around this paper. We would also like to thank the anonymous reviewers and the meta-reviewer for their helpful comments which helped in shaping this paper.

\subsection{Declaration on Use of AI}

We acknowledge the use of Claude (Anthropic) as a copyediting tool to improve spelling, grammar, punctuation, clarity, and concision. Gemini (Google) was also used for web-search.

\bibliography{aaai2026}


\end{document}